\documentstyle[12pt]{article}
\title{Pionium in the elementary particle decays}
\author{Z.~K.~Silagadze \vspace*{5mm}\\
\small\em JINR, Laboratory of Nuclear Problems, 141 980, Dubna and \\
\small\em  Budker Institute of Nuclear Physics, 630 090, Novosibirsk }
\date{}

\begin{document}
\large
\maketitle

\begin{abstract}
 \small\rm Some atomic decays of elementary particles with pionium
($\pi^+\pi^-$-dimesoatom) in the final state are considered. Only for
K-meson atomic decays are the corresponding branching ratios big enough
to make their experimental study realistic. The $O(\alpha)$ order
corrections to the lifetime of pionium are also calculated.
\end{abstract}

\section{Introduction}
 The pionium ($\pi^+\pi^-$-dimesoatom) was firstly considered by Uretsky
and Palfrey \cite{1}, although only after two decades an experimental
study of such Coulomb dimesoatoms, initiated greatly by L.~L.~Nemenov
\cite{2}, became real. Recently about 200 pionium atoms, produced in
a Ta target by 70\,Gev protons, were observed \cite{3}. But an accurate
pionium lifetime measurement is still absent, and it is expected that
this will be done in a more elaborate version of this \cite{4} or
other \cite{5} proposed experiments. From this measurement a model
independent information about pion scattering lengths can be extracted
\cite{1,6,7}.

 Already in \cite{1} it was realized that pionium can be formed in the
elementary particle decays. Among them the $K^+ \to \pi^+ A_{2\pi}$
decay is most promising \cite{8}. The branching ratios of this and
other similar atomic decays are very small, making their experimental
study an elaborate task. In the case of pionium, matters are further
complicated by the fact that it is a very short lived object.

 Let us note that the only atomic decays of mesons, experimentally studied
at yet, are $\pi^0 \to \gamma A_{2e}$ \cite{9} and $K_L \to \nu A_{
\mu \pi}$ \cite{10}. Their measured branching ratios agrees well to what
is expected theoretically \cite{2,11,12}. Even $O(\alpha)$ corrections
for them were calculated \cite{13,14}, although some part of $O(\alpha)$
corrections for $K_L \to \nu A_{\mu \pi}$, originated from the normalization
constant, was omitted in \cite{14}.

 it is expected that several mesonic factories with very high luminosity
will begin to operate in a near future. The study of atomic decays,
even with pionium in a final state, can become real at these factories.
Below we will consider some atomic decays of the elementary particles
which can be interesting in this respect.

 Further information about dimesoatoms, not presented in this article,
can be found in \cite{15,16,17}.

\section{Pionium in the $\eta , \eta \prime$ and K-meson decays}
 In the nonrelativistic approximation pionium state vector coincides
(up to irrelevent phase factor) to the following superposition \cite{12} :
\begin{eqnarray}
\vert P_A;A_{2\pi} \rangle = \int \frac{d\overrightarrow q}{(2\pi)^3}
\, \vert \frac{P_A}{2}+q;\pi^+\rangle \vert \frac{P_A}{2}-q;\pi^-\rangle
\frac{\Psi(\overrightarrow q)}{\sqrt{m}} \hspace*{5mm} ,
\label{eq1}\end{eqnarray}
\noindent
where $\Psi(\overrightarrow q)$ is a momentum space wave function for
$A_{2\pi}$ Coulomb atom and $\sqrt{m}\equiv \sqrt{m_{\pi^+}} \approx
\sqrt{\frac{M_A}{2}}$ appears because of relativistic normalization
\begin{eqnarray}
\langle \overrightarrow{p} \vert \overrightarrow{q} \rangle
=2E_p(2\pi)^3\delta(\overrightarrow{p}-\overrightarrow{q}) \hspace*{5mm},
\hspace*{5mm} E_p=\sqrt{{\overrightarrow p}^2+m^2} \hspace*{5mm},
\label{eq2}\end{eqnarray}
\noindent
which we will use throughout this paper for any one-particle (composed
or elementary) state vector.

 (1) means that an amplitude for the atomic decay
 $P_1 \longrightarrow P_2 + A_{2\pi}$ is given by the following expression
\begin{eqnarray}
\langle P_2A_{2\pi} \vert P_1 \rangle = I(Q_1,\frac{P_A}{2},
\frac{P_A}{2},Q_2) \frac{i \Psi  ( \overrightarrow{x} =0)}
{ \sqrt{m}} \hspace*{15mm}  ,
\label{eq3} \end{eqnarray}
\noindent
where $\Psi (\overrightarrow{x}=0)$ is a Schr\"odinger wave function
for a hydrogenlike atom at the origin, and  $I(Q_1,p_+,p_-,Q_2)$
is an amplitude for the nonatomic decay
$P_1(Q_1) \longrightarrow \pi ^+ (p_+)+\pi^-(p_-)+P_2(Q_2)$.
If this amplitude is known, (3) allows us to calculate a ratio
of atomic and nonatomic decay widths.

 If $P_1$ and $P_2$ are pseudoscalar mesons, then in a $P_1$-meson rest
frame  $I(Q_1,p_+,p_-,Q_2)$ depends only on two independent energies,
and it is convenient to use instead of them the conventional Dalitz
variables:
\begin{eqnarray}
x=\frac{\sqrt{3}(E_+ - E_-)}{M_1-2m-M_2}
\hspace*{10mm} y=3 \frac{M_1-E_+ -E_- - M_2}
{M_1-2m-M_2} - 1 \hspace*{5mm} .
\label{eq4}\end{eqnarray}

 If $M_1-2m-M_2 \ll M_1$, it is expected \cite{18}
that higher order terms in a power series expansion of $I(x,y)$ will
be suppressed and in a good approximation
$$\vert I(x,y) \vert^2 \sim 1+ay+by^2+cx^2 \hspace*{10mm} .$$
\noindent
Note that terms linear in x are forbidden by CP-invariance.
 After integrating over two and three particle phase spaces:
$$\Gamma(P_1 \to P_2A_{2\pi})=$$
$$=\int \vert \langle P_2A_{2\pi} \vert P_1
\rangle \vert^2 \frac{1}{2M_1}(2\pi)^4 \delta (Q_1-Q_2-P_A)
\frac{d \overrightarrow {Q_2}}{(2\pi)^3 2E_2} \frac{d \overrightarrow
P_A}{(2\pi)^3 2E_A}$$
\vspace*{3mm}
$$\Gamma(P_1 \to P_2 \pi^+ \pi^-)= $$
$$=\int \vert I(Q_1,p_+,p_-,Q_2)
\vert^2 \frac{1}{2M_1}(2\pi)^4 \delta (Q_1-Q_2-p_+ -p_-)
\frac{d \overrightarrow {Q_2}\, d \overrightarrow
{p_+}\, d \overrightarrow {p_-}}{(2\pi)^9 8E_2E_+E_-}$$
\noindent
and using $\vert \Psi(\overrightarrow x=0) \vert = \frac{\alpha^3 m^3}
{8\pi}$, we obtain:
\begin{eqnarray}
\frac{\Gamma(P_1 \to P_2 A_{2\pi})}{\Gamma(P_1 \to P_2 \pi^+ \pi^-)}=
\frac{\pi}{R}\alpha^3 (\frac{m}{M_1})^2 \vert I(\tilde x,\tilde y) \vert^2
\sqrt{r^2-\frac{M_2^2}{M_1^2}} \hspace*{5mm}.
\label{eq5}\end{eqnarray}
\noindent
$\tilde x$ and $\tilde y$ are Dalitz variables, which correspond to
the atomic decay:
\begin{eqnarray}
\tilde x=0 \hspace*{5mm} \tilde y=3\frac{\tilde E_2-M_2}{M_1-2m-M_2}
\hspace*{5mm} \tilde E_2=\frac{M_1^2-4m^2+M_2^2}{2M_1}
\hspace*{5mm}. \label{eq6} \end{eqnarray}
R is a dimensionalless remnant of the three-particle phase space integral:
\begin{eqnarray}
R=\int_{(x_+)_{\rm min}}^{(x_+)_{\rm max}} \, dx_+ \, \int_{(x_-)_{\rm
min}}^
{(x_-)_{\rm max}} \, dx_- \, \vert I(x_+,x_-) \vert^2 \hspace*{5mm},
\label{eq7}\end{eqnarray}
\noindent
where $x_+=\frac{E_+}{M_1}, x_-=\frac{E_-}{M_1}$, and the
integration limits in (7) are given by
\begin{eqnarray} &&
(x_+)_{\rm min}=\frac{m}{M_1} \hspace*{20mm}
(x_+)_{\rm max}=\frac{1}{2}(1-\frac{M_2(2m+M_2)}
{M_1^2}) \nonumber \\ &&
\vspace*{6mm}
(x_-)_{\stackrel{min}{max}}=
\frac{1}{2(1-2x_+ + \frac{m^2}{M_1^2})}
\{ (1-x_+)(1-2x_+ + \frac{2m^2-M_2^2}{M_1^2}) \mp
\nonumber \\ && \hspace*{-12mm}
\mp \sqrt{(x_+^2-\frac{m^2}{M_1^2})
(1-2x_+ + \frac{M_2(2m-M_2)}{M_1^2})
(1-2x_+ - \frac{M_2(2m+M_2)}{M_1^2})} \}.
\label{eq8}\end{eqnarray}
At last
\begin{eqnarray}
r=\frac{1}{2}(1-4\frac{m^2}{M_1^2}+\frac{M_2^2}
{M_1^2}) \hspace*{5mm}. \label{eq9} \end{eqnarray}

 The Dalitz-plot distribution for $\eta \to \pi^+ \pi^- \pi^0$ decay
had been measured \cite{19} and the result is
\begin{eqnarray}
\vert I_\eta(x,y) \vert ^2 \sim 1-(1.08\pm0.014)y+
(0.03\pm0.03)y^2+(0.05\pm0.03)x^2
\label{eq10}\end{eqnarray}
Inserting this into (5) and (7), we get (it is assumed that $A_{2\pi}$
is produced in a $1S$ state. If we sum up over all $nS$ states, the
result will increase $\sum_{n=1}^\infty \frac{1}{n^3} \approx 1.2$
times):
\begin{eqnarray}
\frac{\Gamma(\eta \to \pi^0 A_{2\pi})}{\Gamma(\eta \to
\pi^+ \pi^- \pi^0 )} \approx 0.91 \cdot 10^{-7}
\label{eq11}\end{eqnarray}
 Let us note, that a value $3.9 \cdot 10^{-7}$, cited in \cite{16},
corresponds to a theoretical prediction from effective chiral lagrangian
\cite{20} $I(x,y) \sim 1-0.55y$ and seems to be too optimistic, though
the accuracy of quadratic terms determination in (10) allows, in principle,
to increase (11) several times.

 The results, analogous to (10), exist for $K^+ \to \pi^+ \pi^+ \pi^-$
decay \cite{21}:
\begin{eqnarray} &&
\vert I_{K^+}(x,y) \vert ^2 \sim  1+(0.2814\pm0.0082)y-
(0.001\pm0.023)y^2- \nonumber \\ && -(0.099\pm0.019)x^2 \hspace*{5mm},
\nonumber\end{eqnarray}
\noindent
for $K_L \to \pi^+ \pi^- \pi^0$ decay \cite{22}:
\begin{eqnarray} &&
\vert I_{K_L}(x,y) \vert ^2 \sim 1-(0.917\pm0.013)y+
(0.149\pm0.013)y^2+ \nonumber \\ && +(0.055\pm0.010)x^2 \hspace*{5mm},
\nonumber\end{eqnarray}
\noindent
and for $\eta^\prime \to \eta \pi^+ \pi^-$ decay \cite{23}:
\begin{eqnarray}
\vert I_{\eta \prime}(x,y) \vert ^2 \sim
\vert 1-(0.08\pm0.03)y) \vert^2
\hspace*{5mm}.
\nonumber\end{eqnarray}
 Using them, we get
\begin{eqnarray} && \hspace*{33mm}
\frac{\Gamma(K^+ \to \pi^+ A_{2\pi})}{\Gamma(K^+ \to \pi^+ \pi^- \pi^0)}
\approx 10^{-5} \nonumber \\ &&
\frac{\Gamma(K_L \to \pi^0 A_{2\pi})}{\Gamma(K_L \to \pi^+ \pi^- \pi^0)}
\approx 8.6 \cdot 10^{-7} \hspace*{10mm}
\frac{\Gamma(\eta^\prime \to \eta A_{2\pi})}
{\Gamma(\eta^\prime \to \eta \pi^+ \pi^- )}
\approx 1.4 \cdot 10^{-6}
\label{eq12}\end{eqnarray}
 For $K^+ \to \pi^+A_{2\pi}$ decay, it is necessary to take into account
the identity of $\pi^+$-mesons, which increases the result two times.
An extra $\sim 5$ times difference between $K^+$ and $K_L$ decays is
due to $I_{K^+}(\tilde x,\tilde y)/I_{K_L}(\tilde x,\tilde y) \approx 2.15$.
Note that $K^+ \to \pi^+A_{2\pi}$ decay was considered earlier in \cite{8}
with slightly different result.

 Taking nonatomic decays branching ratios from \cite{24}, (11) and (12)
can be transformed to
\begin{eqnarray} && \hspace*{-10mm}
Br(\eta \to \pi^0 A_{2\pi}) \approx 2 \cdot 10^{-8} \hspace*{19mm}
Br(\eta^\prime \to \eta A_{2\pi}) \approx 6.2 \cdot 10^{-7} \nonumber \\ &&
\hspace*{-10mm} Br(K^+ \to \pi^+ A_{2\pi}) \approx 5.5 \cdot 10^{-7}
\hspace*{10mm} Br(K_L \to \pi^0 A_{2\pi}) \approx 1.1 \cdot 10^{-7}
\label{eq13}\end{eqnarray}

 For $\phi$-factory (13) means about $10^4$ K-meson atomic decays
with pionium per year. So, we think, the study of such atomic decays
at $\phi$-factory is not only realistic, but a desirable task.

\section{Pionium in the $\psi$ and $\Upsilon$-meson decays}
 For $c-\tau$ and B-factories $\psi(2S) \to \psi(1S) A_{2\pi}$ and
$\Upsilon(2S) \to \Upsilon(1S) A_{2\pi}$ decays can be interesting since
the corresponding nonatomic decays have large branching ratios.

 In the nonrelativistic approximation the most general form for the
$V_1 \to V_2 \pi^+ \pi^-$ decay amplitude, which follows from PCAC,
is \cite{25}
\begin{eqnarray} &&
\langle V_2\pi^+ \pi^- \vert V_1 \rangle =
\vec{\epsilon_1} \cdot
\vec{\epsilon_2}[-A\, q_+\cdot q_- + B\, E_+ E_-]+ \nonumber \\ &&
+C\, (\vec{\epsilon_1}\cdot \vec{q_1} \, \vec{\epsilon_2}\cdot \vec{q_2}+
\vec{\epsilon_1}\cdot \vec{q_2} \, \vec{\epsilon_2}\cdot \vec{q_1})
\label{eq14}\end{eqnarray}
\noindent
where A,B,C are approximately constant and $\epsilon_1,\epsilon_2$
are $V_1,V_2$ vector meson polarization vectors.

 There are some theoretical indications \cite{26} and experiment
confirms \cite{27} that $C=0$. Then to calculate the (5) ratio,
we only need B/A ratio and it can be extracted from the $\pi^+ \pi^-$
invariant mass distribution in the $V_1 \to V_2\pi^+\pi^-$ decay.
The results are \cite{27}:
\begin{eqnarray}
\frac{B}{A}=-0.21\pm0.01
\label{eq15}\end{eqnarray}
\noindent
for the $\psi(2S) \to \psi(1S) A_{2\pi}$ decay, and
\begin{eqnarray}
\frac{B}{A}=-0.154\pm0.019
\label{eq16}\end{eqnarray}
\noindent
for the $\Upsilon(2S) \to \Upsilon(1S) A_{2\pi}$ one.

 After summing over vector meson polarizations, it follows from (14) that
(if C=0)
\begin{eqnarray}
\langle V_2\pi^+ \pi^- \vert V_1 \rangle \sim \frac{1}{2}(1+
\frac{M_2^2}{M_1^2}-2\frac{m^2}{M_1^2})-1+x_+ + x_- -\frac{B}{A}x_+ x_-
\hspace*{5mm},
\label{eq17}\end{eqnarray}
\noindent
$x_+$ and $x_-$ were defined earlier.

 Using this instead of $I(x,y)$ in (5) and (7), we get from (15) and (16)
\begin{eqnarray} &&
\frac{\Gamma(\psi(2S) \to \psi(1S) A_{2\pi})}
{\Gamma(\psi(2S) \to \psi(1S) \pi^+ \pi^-)}
\approx 4.6 \cdot 10^{-8} \nonumber \\ &&
\vspace*{5mm}
\frac{\Gamma(\Upsilon(2S) \to \Upsilon(1S) A_{2\pi})}
{\Gamma(\Upsilon(2S) \to \Upsilon(1S) \pi^+ \pi^- )}
\approx 5.2 \cdot 10^{-8} \hspace*{5mm},
\label{eq18}\end{eqnarray}
\noindent
which correspond to the following branching ratios
\begin{eqnarray} &&
Br(\psi(2S) \to \psi(1S) A_{2\pi}) \approx 1.4 \cdot 10^{-8} \nonumber \\ &&
Br(\Upsilon(2S) \to \Upsilon(1S) A_{2\pi}) \approx 10^{-8}
\label{eq19}\end{eqnarray}
 Unfortunately this is too small for B-factory. We can expect only
several events per year. So it seems unrealistic to study $\Upsilon$-meson
atomic decays at B-factory.

\section{$O(\alpha)$ order corrections to the pionium lifetime}
 The main decay mode for pionium is $A_{2\pi} \to \pi^0 \pi^0$ and
its amplitude according to Mandelstam \cite{28} can be expressed in the
form
\begin{eqnarray}
\langle \pi^0 \pi^0 \vert A_{2\pi} \rangle =
\int \frac{dp}{(2\pi)^4} J(p_1,p_2,\frac{P_A}{2}+p,
\frac{P_A}{2}-p) \chi (p) \hspace*{10mm},
\label{eq20}\end{eqnarray}
\noindent
where $\chi(p)$ is $A_{2\pi}$-dimesoatom bound state Bethe-Salpeter wave
function and $J(p_1,p_2,p_+,p_-)$ stands for
$\pi^+ \pi^-$-irreducible kernel for the reaction \linebreak
$\pi^+(p_+)+\pi^-(p_-) \to \pi^0(p_1)+\pi^0(p_2)$ .

Up to $O(\alpha)$ terms, J is constant defined through pion scattering
lengths $a_0$ and $a_2$ \cite{29}:
\begin{eqnarray}
J=\frac{32}{3} \pi m(a_0-a_2) \hspace*{15mm}.
\label{eq21}\end{eqnarray}
\noindent
So (20) can be rewritten as
\begin{eqnarray}
\langle \pi^0 \pi^0 \vert A_{2\pi} \rangle =
J \cdot \chi(x=0) \hspace*{15mm},
\label{eq22}\end{eqnarray}
\noindent
where
\begin{eqnarray}
\chi(x=0)=\int \frac{dp}{(2\pi)^4} \chi (p)
\label{eq23}\end{eqnarray}
\noindent
is a configuration space Bethe-Salpeter wave function at the origin.

 The Bethe-Salpeter equation for $\chi(p)$ in a $A_{2\pi}$ rest frame,
up to $O(\alpha)$ terms, takes the form
\begin{eqnarray} &&
\left[m^2+\overrightarrow{p}^2-(\frac{M_A}{2}+p_0)^2\right]
\left[m^2+\overrightarrow{p}^2-(\frac{M_A}{2}-p_0)^2\right]\chi(p)= \nonumber
\\ &&
=\frac{i\lambda}{\pi^2}\int dq \, \frac{\chi(q)}{(p-q)^2-i\epsilon}
\label{eq24}\end{eqnarray}
\noindent
where $\lambda= \frac{\alpha M_A^2}{4\pi}$.

 (24) corresponds to the Wick-Cutkosky model \cite{30}. Let us note that
this fact was firstly noticed and used to calculate $O(\alpha)$ order
corrections to the $K_L \to \nu A_{\mu \pi}$ decay width in \cite{14}.

 According to \cite{30} (see also \cite{31} for a review), a ground state
($1S$ in the nonrelativistic limit) solution of (24) corresponds to
\begin{eqnarray}
\chi(p)=\int_{-1}^1 \frac {g(z) \,dz}{[A+Bz]^3}
\hspace*{15mm},
\label{eq25}\end{eqnarray}
\noindent
where
\begin{eqnarray}
A=m^2-\frac{1}{4}M_A^2-p^2 \equiv \Delta^2-p^2
\hspace*{10mm} B=p_0M_A \hspace*{5mm},
\nonumber\end{eqnarray}
\noindent
and g(z) spectral function satisfies the following integral equation
\begin{eqnarray}
g(z)=\frac{\lambda}{2} \int_{-1}^1 \frac{1}{\Delta^2+
\frac{1}{4}M_A^2y^2}[\frac{1-z}{1-y}\Theta (z-y) +
\frac{1+z}{1+y} \Theta (y-z)]g(y) \,dy
\label{eq26}\end{eqnarray}
\noindent
If $M_A=2m-E,\, E\ll m$, then $\Delta^2 \approx mE$ and
\begin{eqnarray}
\frac{1}{\Delta^2+\frac{1}{4}M_A^2y^2}=\frac{\pi}{m\sqrt{mE}}
\delta (y)  \hspace*{15mm},
\label{eq27}\end{eqnarray}
\noindent
because $$\frac{\epsilon}{\epsilon^2+y^2}
\longrightarrow \pi \delta(y) \hspace*{5mm}.$$
\noindent
when $\epsilon \to 0$.

Substituting (27) into (26), we get
\begin{eqnarray}
g(z)=\frac{\lambda \pi}{2m\sqrt{mE}}(1-\vert z \vert)\, g(0)
\hspace*{10mm}.
\label{eq28}\end{eqnarray}
\noindent
So
\begin{eqnarray}
\frac{\lambda \pi}{2m\sqrt{mE}}=1 \hspace*{10mm},
\nonumber\end{eqnarray}
\noindent
which really gives a hydrogenlike atom nonrelativistic ground state
energy level $E=\frac{m\alpha^2}{4}$.

 But the solution (28), found in \cite{30}, is not a complete $O(\alpha)$
order solution. Indeed, taking $g(z)=g_0(z)+\alpha g_1(z)$ and
\begin{eqnarray}
\frac{1}{\Delta^2+\frac{1}{4}M_A^2y^2}=\frac{2\pi}{m^2\alpha}
\delta (y) + \sigma(y) \hspace*{15mm},
\label{eq29}\end{eqnarray}
\noindent
where $\sigma (y)$ has $O(\alpha)$ order smallness compared to the first
$\sim \delta (y)$ term, we get from (26) ($N_0$ and $N_1$ are constants)
\begin{eqnarray} \hspace*{-2mm}
g_0(z)=N_0(1-\vert z \vert ) \hspace*{5mm} g_1(z)=N_1(1-\vert z \vert)+
\frac{\lambda}{2\alpha}\int_{-1}^1 \sigma (y) R(z,y)g_0(y) \,dy
\label{eq30}\end{eqnarray}
\noindent
where
$$ R(z,y)=\frac{1-z}{1-y}\Theta (z-y) +
\frac{1+z}{1+y} \Theta (y-z) \hspace*{10mm}. $$
Calculating in the $\alpha \to 0$ limit the integral in (30), we get
$$ g_1(z)=N_1(1- \vert z \vert) +\frac{N_0}{\pi} \{ (1-
\vert z \vert )\ln(\alpha)+(1+ \vert z \vert)[\ln(2 \vert z \vert )-
\ln(1+ \vert z \vert)] \} \hspace*{3mm} . $$
Therefore, the complete $O(\alpha)$ order solution of (26) looks like
\begin{eqnarray}
g(z)=N\{ (1- \vert z \vert )+ \frac{\alpha}{\pi}(1+ \vert z \vert)
[\ln(2\vert z \vert)-\ln(1+ \vert z \vert ) ] \} \hspace*{10mm}
\label{eq31} \end{eqnarray}
Substituting this in (25), we get the $O(\alpha)$ order pionium
Bethe-Salpeter wave function
\begin{eqnarray} &&
\chi (p,P_A)= \nonumber \\ &&
=\frac{N}{(\Delta^2-p^2)\left[m^2-(\frac{P_A}{2}+p)^2\right]
\left[m^2-(\frac{P_A}{2}-p)^2\right]}\left\{ 1+ \frac{\alpha}{\pi}
\chi_1(p,P_A) \right\} \hspace*{5mm},
\label{eq32}\end{eqnarray}
\noindent
where
\begin{eqnarray} &&
\chi_1(p,P_A)=\frac{m^2-(\frac{P_A}{2}-p)^2}{2(\Delta^2-p^2)}
\ln(m^2-(\frac{P_A}{2}-p)^2)+ \nonumber \\ && \hspace*{-5mm}
+\frac{m^2-(\frac{P_A}{2}+p)^2}{2(\Delta^2
-p^2)}\ln(m^2-(\frac{P_A}{2}+p)^2)
-\ln(\Delta^2-p^2)+O(\alpha \ln(\alpha))
\label{eq33}\end{eqnarray}

 A normalization constant N is defined from the normalization condition
\cite{32}, which in the $O(\alpha)$ order takes the form ($A_{2\pi}$ rest
frame is assumed)
\begin{eqnarray}  &&
-2m^2\alpha^2 \approx \alpha \frac{dM^2}{d\alpha} = \nonumber \\ &&
\hspace*{-10mm} = \frac{iN^2}{(2\pi)^4} \int \frac{dp}{(\Delta^2-p^2)^2
\left[ \left( \frac{M_A}{2}+p_0 \right)^2-m^2-\vec{p}^2 \right]
\left[ \left( \frac{M_A}{2}-p_0 \right)^2-m^2-\vec{p}^2 \right]}
\label{eq34}\end{eqnarray}
\noindent
But the integral in the r.h.s. of (34) equals to
\begin{eqnarray} &&
L=- \frac{i\pi}{2} \frac{\partial^2}{\partial (\Delta^2)^2 } \int d \vec{p}
\int \frac{dp_0}{2\pi i} \frac{1}{\Delta^2+\vec{p}^2-p_0^2 }
[\frac{1}{(\frac{M_A}{2}+p_0)^2-m^2-\vec{p}^2 }+ \nonumber \\ &&
+\frac{1}{(\frac{M_A}{2}-p_0)^2-m^2-\vec{p}^2 }]
\nonumber \end{eqnarray}
 Remembering that in fact $m^2$ in the above expression should be replaced
by $m^2-i\epsilon$, we can perform an integration over $dp_0$
and obtain
\begin{eqnarray} && \hspace*{-5mm}
L=\frac{2i\pi^2}{M_A^2}\int_0^\infty dx \, \frac{x^2}{(\Delta^2+x^2)^2}
\{ \frac{1}{\sqrt{m^2+x^2}}(\frac{4(m^2+x^2)}{\Delta^2+x^2}-1)
-\frac{3}{\sqrt{\Delta^2+x^2}} \} \approx \nonumber \\ && \hspace*{-5mm}
\approx \frac{i\pi^3}{m^4\alpha^3}(1-2\frac{\alpha}{\pi}+O(\alpha^2))
\nonumber \end{eqnarray}
\noindent
So (34) takes the form
$$-2m^2\alpha^2=\frac{i}{(2\pi)^4}N^2\frac{i\pi^3}{m^4\alpha^3}
(1-2\frac{\alpha}{\pi}) \hspace*{10mm},$$
\noindent
and therefore
\begin{eqnarray}
N=32\sqrt{\pi m}(\frac{1}{2}m\alpha)^{5/2}
(1+\frac{\alpha}{\pi}) \hspace*{15mm}.
\label{eq35}\end{eqnarray}
 Analogously
\begin{eqnarray} &&
\int \frac{dp}{(2\pi)^4}\, \chi(p)=
\frac{iN}{8\pi^2M_A^2}\int_0^\infty dx \, \frac{x^2}{\Delta^2+x^2}
\{ \frac{2}{\sqrt{m^2+x^2}} - \frac{2}{\sqrt{\Delta^2+x^2}}+ \nonumber \\ &&
+\frac{M_A^2}{\sqrt{(m^2+x^2)(\Delta^2+x^2)}} \} \approx
\frac{iN}{16\pi m^2\alpha}(1+\frac{\alpha}{\pi})
\hspace*{10mm}. \nonumber \end{eqnarray}
\noindent
Therefore
\begin{eqnarray}
\chi(x=0)=\frac{iN}{32m\pi(\frac{1}{2}m\alpha)}(1+\frac{\alpha}{\pi})
\approx \frac{i}{\sqrt{m}}\Psi(\vec{x}=0)\, (1+2\frac{\alpha}{\pi})
\label{eq36}\end{eqnarray}
Substituting this in (22), we finally get
\begin{eqnarray}
\Gamma(A_{2\pi} \to \pi^0 \pi^0)=\Gamma_0 (A_{2\pi} \to \pi^0 \pi^0)
(1+4\frac{\alpha}{\pi}) \hspace*{15mm},
\label{eq37}\end{eqnarray}
\noindent
where \cite{6}
$$ \Gamma_0 (A_{2\pi} \to \pi^0 \pi^0)=\frac{16\pi}{9}
(a_0-a_2)^2\sqrt{\frac{2(m_{\pi^+}-m_{\pi^0})}{m_{\pi^+}}}
\vert \Psi(\vec{x}=0) \vert^2 \hspace*{5mm}. $$

$\sim \chi_1$ part of the Bethe-Salpeter wave function  contributes
neither in its value at the origin nor in its normalization, because
in the $O(\alpha)$ order
$$\int \frac{\alpha \, dp_0}
{\left[ \left( \frac{M_A}{2}+p_0 \right)^2-m^2-\vec{p}^2 \right]
\left[ \left( \frac{M_A}{2}-p_0 \right)^2-m^2-\vec{p}^2 \right]}
\cdots =\int \frac{i\pi \alpha \delta(p_0) \, dp_0}{M_A(\Delta^2+\vec{p}^2)}
\cdots $$
\noindent
and $\delta(p_0)\chi_1(p;P_A)=0$.

 As a last remark, let us note that a nonrelativistic approximation (3)
for the $M_1 \to M_2 + A_{2\pi}$ decay amplitude follows from
$$\chi(x=0) \approx \frac{i}{\sqrt{m}}\Psi(\vec{x}=0)$$
\noindent
and \cite{28}
\begin{eqnarray} &&
\langle P_2A_{2\pi} \vert P_1 \rangle = \int \frac{dp}{(2\pi)^4}
I(Q_1,\frac{P_A}{2}+p,\frac{P_A}{2}-p,Q_2) \chi(p) \approx \nonumber \\ &&
\approx I(Q_1,\frac{P_A}{2},\frac{P_A}{2},Q_2)\chi(x=0) \hspace*{3mm}.
\nonumber \end{eqnarray}

\vspace*{10mm}
 Author is grateful to A.~M.~Khvedelidze for helpful discussions and
critical remarks and also to N.~V.~Makhaldiani, E.~A.~Kuraev,
V.~N.~Pervushin and L.~K.~Lytkyn for their interest in this work.

\vspace*{5mm}
email (till March,1995):  silagadze@lnpvx2.jinr.dubna.su

\end{document}